\documentclass{llncs}
\usepackage{graphicx}
\usepackage{multirow}
\usepackage{mathtools}
\usepackage{diagbox}
\begin{document}
\title{On Applying Meta-path for Network Embedding in Mining Heterogeneous DBLP Network}

%
%
\author{Akash Anil\and
Uppinder Chugh  \and
Sanasam Ranbir Singh}
%
%

\institute{Department of Computer Science and Engineering \\
	Indian Institute of Technology Guwahati \\
\email{<a.anil.iitg, uppinderchugh>@gmail.com, ranbir@iitg.ac.in}}

\maketitle              
\begin{abstract}
In recent time, applications of network embedding in mining real-world information network have been widely reported in the literature. Majority of the information networks are heterogeneous in nature.
Meta-path is one of the popularly used approaches for generating embedding in heterogeneous networks. As meta-path guides the models towards a specific sub-structure, it tends to lose some heterogeneous characteristics inherently present in the underlying network. In this paper, we systematically study the effects of different meta-paths using different state-of-art network embedding methods (Metapath2vec, Node2vec,
and VERSE) over DBLP bibliographic network and evaluate the performance of embeddings using two applications (co-authorship prediction
and author’s research area classification tasks). From various experimental observations, it is evident that embedding using different meta-paths
perform differently over different tasks. It shows that meta-paths are
task-dependent and can not be generalized for different tasks. We further observe that embedding obtained after considering all the node and
relation types in bibliographic network outperforms its meta-path based
counterparts.

\end{abstract}

\section{Introduction}
\label{sec:introduction}
Recently there is a surge in applying network embedding for addressing various tasks in network science such as classification, clustering, link prediction, community detection etc.~\cite{Grover:2016:NSF:2939672.2939754,dong2017metapath2vec,perozzi2014deepwalk,Tsitsulin:2018:VVG:3178876.3186120}. Network embedding aims at learning low dimensional feature vector for a node which is capable of preserving its structural characteristics~\cite{cao2015grarep,Grover:2016:NSF:2939672.2939754}. Majority of network embedding models proposed in the past focus mainly on mining homogeneous networks consisting of singular type of node and relation~\cite{dong2017metapath2vec}. However, many of the real-world information networks and social networks are heterogeneous in nature consisting of different types of nodes and relations
~\cite{sun2012mining}. For example, an academic bibliographic network may be better represented using author, paper, venue (conference/journal) as nodes and different contextual relations such as author-writes-paper, author-publishes-at-venue, etc.

Majority of the previous studies on mining heterogeneous networks~\cite{cao2014collective,sun2011co} exploit {\it meta-paths}~\cite{kong2012meta} which is a sequence of relations between different node types as defined below. Given a heterogeneous network $G(N,E,\mathcal{V}, \mathcal{R})$ where $N$, $E$, $\mathcal{V}$, and $\mathcal{R}$ are set of nodes, set of relations, set of node types, and set of relation types, a meta-path 
$\mathcal{P}$ is defined as a path
$\mathcal{P}:v_1\xrightarrow{r_1}v_2\xrightarrow{r_2}v_3\xrightarrow{r_{3}}...$ where $v_i\in \mathcal{V}$ and $r_i\in\mathcal{R}$. For example, a meta-path $author\xrightarrow{writes}paper\xrightarrow{writtenBy}author$ in a bibliographic network represents a co-authorship relation in a paper. While exploring a network, a meta-path defines the specific path the explorer should follow.  In the past, meta-paths have been used to generate network embedding~\cite{dong2017metapath2vec} and reported to obtain promising results for various applications. In this paper, we systematically analyze effectiveness of considering meta-path for generating network embedding, specifically for bibliographic network. Since, meta-path guides to explore only the partial network defined by the meta-path, it may lose some of the inherent network properties. Motivated by this, this paper attempts to understand the following two important issues while considering meta-paths for generating network embedding. 
\begin{enumerate}	
	\item Does meta-path lose network information which can degrade the network embedding performance? 
	\item Are meta-path based embeddings independent to the end task?

\end{enumerate}	

To investigate the above-discussed problems, we evaluate embeddings generated using different types of meta-paths using three state-of-art embedding models namely, (i) Metapath2vec~\cite{dong2017metapath2vec}, (ii) Node2vec~\cite{Grover:2016:NSF:2939672.2939754}, and (iii) VERSE~\cite{Tsitsulin:2018:VVG:3178876.3186120} on co-authorship prediction task and author's research area classification in DBLP\footnote{https://dblp.uni-trier.de/} heterogeneous bibliographic network. From various experimental observations, it is evident that embeddings generated using entire heterogeneous network outperform the embedding generated using specific meta-paths. Further, It is also observed that 
embedding using different meta-paths may perform differently over different tasks if not chosen carefully.

Rest of the paper is organized as follows. Section~\ref{sec:literature_survey} reviews some of the previous works on network embedding. Section~\ref{sec:experimental_analysis} describes the experimental setups and results. Paper concludes in Section ~\ref{sec:conclusion}.

\section{Literature Survey}
\label{sec:literature_survey}
For network embedding, a majority of the initial studies attempt to map the natural graph representations like normalized adjacency or Laplacian matrix to lower dimensions by using spectral graph theory~\cite{belkin2002laplacian,ou2016asymmetric} and various non-linear dimensionality reduction techniques~\cite{roweis2000nonlinear,tenenbaum2000global,ahmed2013distributed}. However, these models are not scalable to large real world networks as they exploit graph decomposition techniques at the core which requires whole matrix beforehand.

To overcome the above limitations, many network embedding models exploit a framework which first generates a neighborhood sample using a random walk or proximity measure and then leverages it to learn the node embeddings using a skip-gram~\cite{mikolov2013distributed} based neural network model~\cite{Grover:2016:NSF:2939672.2939754,perozzi2014deepwalk,tang2015line}. For example, Node2vec~\cite{Grover:2016:NSF:2939672.2939754} uses a $2nd$ order random walk to generate the sample and learn the node embedding using skip-gram model. Further, VERSE~\cite{Tsitsulin:2018:VVG:3178876.3186120} preserves the vertex-to-vertex similarity using personalized PageRank and thereby uses a single layer neural network to learn the embeddings.     

All the above graph embedding models were proposed for homogeneous network. Recently, Metapath2vec~\cite{dong2017metapath2vec} first proposes embedding model for heterogeneous networks which samples the node neighborhood using a random walk guided by meta-path, and then uses skip-gram model to learn the node embedding.       

\begin{table*}[t]
\centering
\caption{Characteristics of different networks constructed over DBLP data}
\label{tab:data_char}
\resizebox{\textwidth}{!}{
\begin{tabular}{|c|c|cc|cc|ccc|c|}
\hline
    & \multicolumn{8}{c|}{\textbf{DBLP 1968-2008}}                                                                                              & \textbf{DBLP 2009-2011}                                                                  \\ \cline{2-9}
\textbf{Dataset}  & \multicolumn{1}{c|}{\textbf{AA}}      & \multicolumn{2}{c|}{\textbf{APA}} & \multicolumn{2}{c|}{\textbf{AVA}} & \multicolumn{3}{c|}{\textbf{All}}            &  \\ \hline
\textbf{Node Types} & Author & Author    & Paper    & Author    & Venue    & Author & Paper & Venue & Author                                                                   \\ \hline
\textbf{\# Nodes} & 162298  & 162298             & 155189            & 162298             & 621               & 162298          & 155189         & 621            & 18457                                                                             \\ \hline
\textbf{\#Edges} & \multicolumn{1}{c|}{461722}   & \multicolumn{2}{c|}{475828}            & \multicolumn{2}{c|}{326602}            & \multicolumn{3}{c|}{957856}                       & 29677                                                                             \\ \hline
\end{tabular}
}
\end{table*}

\section{Experimental Setups and Analysis}
\label{sec:experimental_analysis}

\subsection{Experimental Dataset}
\label{subsec:dataset}
This paper uses DBLP bibliographic dataset (reported in~\cite{yang2012topic}) covering publication information for the period between 1968 to 2011. To generate various network embeddings using different meta-paths and evaluate the embedding performance over different applications, we further divide the dataset into two parts; (i) between 1968 to 2008 for generating network embedding, and (ii) between 2009 to 2011 for evaluating the embedding over different applications.
This paper considers three types of heterogeneous entity classes namely (i) Author (A), (ii) Paper (P), (iii) Venue (V) for constructing various classes of networks defined by different meta-paths.
 We construct the following four different types of undirected networks from the DBLP 1968-2008 dataset.
\begin{itemize}
\item \texttt{AA}: It is a homogeneous unweighted co-authorship network considering only {\tt Author} node type. Two nodes are connected if they co-author a paper.
\item \texttt{APA}: It is a heterogeneous unweighted network considering {\tt Author} and {\tt Paper} node classes. An author is connected to a paper if he/she is one of the authors of the paper. 
\item \texttt{AVA}: It is a heterogeneous unweighted network considering {\tt Author} and {\tt venue} node classes. An author is connected to a venue if he/she published a paper in that venue. This network structure is similar to the structure considered in {\tt Metapath2vec}~\cite{dong2017metapath2vec}.
\item \texttt{All}: It is a heterogeneous unweighted network  considering all three types of nodes ({\tt Author, Paper}, and {\tt Venue}) and corresponding relationships between them.
\end{itemize}
Table~\ref{tab:data_char} shows the characteristics of these experimental networks.

\subsection{Experimental Setups}
\label{subsec:experimental_setups}
 As mentioned above, three popular recently proposed network embedding models namely (i) Metapath2vec~\cite{dong2017metapath2vec}, (ii) Node2vec~\cite{Grover:2016:NSF:2939672.2939754}, and (iii) VERSE~\cite{Tsitsulin:2018:VVG:3178876.3186120} are considered to generate different node embeddings. For all these models, we use the same hyper-parameter values as described in the original studies cited above. All the embedding results reported in this paper consider 100 dimensional vector~\footnote{While testing with different dimensions 100, 200, 300, we did not observe significant differences. We therefore consider 100 dimensional vector.}. To investigate the performance of different meta-paths and their associated embedding, we evaluate the embedding quality using the following two applications.

\subsubsection{Co-authorship Prediction:}
\label{sec:link_prediction}
Like the study~\cite{Tsitsulin:2018:VVG:3178876.3186120}, we also consider co-authorship prediction task as a classification problem i.e., given a node pair, classify if the node pair has a co-author relation or not. To model it as a binary classification problem, we generate feature vectors representing node pairs using Hadamard operator~\cite{Grover:2016:NSF:2939672.2939754,Tsitsulin:2018:VVG:3178876.3186120}.  
To avoid possible bias with the embedding towards the target application, we consider the DBLP 2009-2011 (non-overlapping with the embedding dataset) for generating samples for the classification task. In this sample, there are 29,677 number of co-authorship relations and 18,457 authors. We use random 80-20 split as training and test samples subjected to 
four different classifiers namely Gaussian Naive Bayes (NB), Random Forest (RF), Decision Tree (DT), and Logistic Regression (LR). To avoid over-fitting, above setup has been repeated 10 times.

\subsubsection{Research Area Classification:} 
We now investigate quality of the embeddings for predicting  author's research area. For each author in DBLP 2009-2011, we further identify (considering the {\tt Field} attribute in~\cite{yang2012topic}) the area in which the author has the maximum publication and consider it as the author's class label. Like co-authorship prediction, we use similar random 80-20 split for all the classifiers and repeated 10 times.

\subsection{Result and Discussion}
\label{subsec:result_and_discussion}
From Tables~\ref{tab:link_pred} and~\ref{tab:classification}, it is observed that LR out-performs other classifiers in 93\% times for co-authorship prediction and 75\% times for research area classification task. Therefore, we select LR Accuracy for further analysis.   

\begin{table*}[t]
\caption{Co-authorship Prediction by Classifiers for different Networks}
\label{tab:link_pred}
\resizebox{\textwidth}{!}{
\begin{tabular}{|c|cccc|cccc|cccc|cccc|}
\hline
                             & \multicolumn{4}{c|}{\textbf{Metapath2vec}}                                   & \multicolumn{4}{c|}{\textbf{Node2vec}}                                       & \multicolumn{4}{c|}{\textbf{VERSE}}
                             & \multicolumn{4}{c|}{\textbf{Combine}}\\ \cline{2-17}
  \textbf{Classifier}                           & \textbf{AA} & \textbf{APA} & \textbf{AVA} & \textbf{All} & \textbf{AA} & \textbf{APA} & \textbf{AVA} & \textbf{All} & \textbf{AA} & \textbf{APA} & \textbf{AVA} & \textbf{All} & \textbf{AA} & \textbf{APA} & \textbf{AVA} & \textbf{All} \\ \hline
\textbf{NB}         & 0.585            & 0.633             & 0.694             & 0.717    & 0.688            & 0.699             & 0.697             & 0.719    & 0.725            & 0.756    & 0.733             & 0.746&   0.673&	0.745&	0.737&	0.758
     \\ \hline
\textbf{RF}       & 0.761   & 0.724             & 0.698             & 0.720             & 0.749   & 0.731             & 0.698             & 0.730             & 0.760   & 0.754             & 0.707             & 0.744 & 0.772&	0.753&	0.714&	0.748
           \\ \hline
\textbf{DT}       & 0.683   & 0.654             & 0.628             & 0.644             & 0.678   & 0.658             & 0.632             & 0.657             & 0.688   & 0.674             & 0.642             & 0.678   &  0.699&	0.673&	0.645&	0.678
         \\ \hline
\textbf{LR} & 0.736            & 0.739             & 0.738             & \textbf{0.766}    & 0.773            & 0.766             & 0.75              & \textbf{0.777}    & 0.788            & 0.784             & 0.764             & \textbf{0.796} &0.799&	0.795&	0.778&	\textbf{0.806}
   \\ \hline
\end{tabular}
}
\end{table*}
\begin{table*}[]
\caption{Author's Research Area Prediction by Classifiers for different Networks}
\label{tab:classification}
\resizebox{\textwidth}{!}{
\begin{tabular}{|c|cccc|cccc|cccc|cccc|}
\hline
                             & \multicolumn{4}{c|}{\textbf{Metapath2vec}}                                   & \multicolumn{4}{c|}{\textbf{Node2vec}}                                       & \multicolumn{4}{c|}{\textbf{VERSE}} & \multicolumn{4}{c|}{\textbf{Combine}}                                         \\ \cline{2-17}
   \textbf{Classifier}
                             & \textbf{AA} & \textbf{APA} & \textbf{AVA} & \textbf{All} & \textbf{AA} & \textbf{APA} & \textbf{AVA} & \textbf{All} & \textbf{AA} & \textbf{APA} & \textbf{AVA} & \textbf{All}& \textbf{AA} & \textbf{APA} & \textbf{AVA} & \textbf{All} \\ \hline
\textbf{NB}         & 0.392            & 0.476             & 0.503    & 0.499             & 0.500            & 0.582    & 0.497             & 0.488             & 0.492            & 0.557    & 0.550             & 0.552        &0.429&	0.58&	0.529&	0.522
     \\ \hline
\textbf{RF}       & 0.484            & 0.486             & 0.491    & 0.482             & 0.488            &0.536    & 0.518             & 0.509             & 0.495            & 0.499             & 0.530             & 0.545  &0.499&	0.529&	0.527&	0.53
 \\ \hline
\textbf{DT}       & 0.442   & 0.439             & 0.439             & 0.428             & 0.436            & 0.481    & 0.472             & 0.449             & 0.445            & 0.440             & 0.476             & 0.490  &0.456&	0.471&	0.474&	0.495
 \\ \hline
\textbf{LR} & 0.504            & 0.539             & 0.565             & \textbf{0.566}    & 0.486            & 0.544             & \textbf{0.559}    & 0.555             & 0.536            & 0.531             & 0.605             & \textbf{0.624}  &0.552&	0.592&	0.612&	\textbf{0.625}
  \\ \hline
\end{tabular}
}
\end{table*}

We first investigate if meta-path based embedding loses 
information or not. 
Tables~\ref{tab:link_pred} and~\ref{tab:classification} present the Accuracy for co-authorship prediction and author's research area classification using three network embedding models discussed above for all networks, i.e. \texttt{AA, AVA, APA}, and \texttt{All}. 

It is evident from Tables~\ref{tab:link_pred} and~\ref{tab:classification} that almost all the models perform best by exploiting All and show poor performance with \texttt{AA, APA} and \texttt{AVA} networks for both tasks, i.e. co-authorship prediction and area classification. Thus, it can be inferred that meta-path alone may be a weak representation for the network because it does not incorporate the impacts of other relational properties while capturing node neighborhood. 

Secondly, we intent to investigate if same embedding responds coherently to different problems. 
From Tables~\ref{tab:link_pred} and~\ref{tab:classification}, it is clearly visible that \texttt{APA} performs better than \texttt{AVA} for co-authorship prediction whereas \texttt{AVA} performs better than \texttt{APA} for classifying author's research area. This observation is true for all the embedding techniques used in this study. Therefore, meta-path based approaches may fail in capturing heterogeneous characteristics of the underlying heterogeneous network if chosen independent to the end task.

Among all the embedding models, VERSE consistently outperforms others for almost all the networks and classifiers for both co-authorship prediction and research area classification tasks. 
     
We further investigate combining all the three embeddings (Metapath2vec, Node2vec, VERSE) by concatenating the feature vectors. From Tables~\ref{tab:link_pred} and~\ref{tab:classification}, it is observed that combined embeddings always out-performs individual embedding for co-authorship prediction and research area classification over all the four networks. 

\section{Conclusion}
\label{sec:conclusion}
In this paper, we investigate the applicability of meta-paths in network embedding 
for co-authorship prediction and author's research area classification problems in heterogeneous DBLP database.
From various experimental results, we observe that by using the entire network majority of the embedding methods out-perform their counter-parts exploiting  meta-path based network for both of the above-discussed tasks. Further,
it is also evident that exploiting past co-authorship relation or \texttt{APA} meta-path yield better co-author prediction in comparison to \texttt{AVA} meta-path which exploits author's publication venue. On the other hand \texttt{AVA} meta-path contributes positively for author's research area classification problem and have superior performance than \texttt{APA} meta-path. 
Thus, for heterogeneous network embedding one should carefully choose the node types, relation types and meta-paths which can capture better the network characteristics to address the underlying problem.

\bibliographystyle{splncs04} 
\bibliography{ref_ecir}

\end{document}